\documentclass[aps,groupedaddress,superscriptaddress,amsmath,amssymb,pra,twocolumn,longbibliography]{revtex4-2}
\usepackage{bm}
\usepackage{hyperref}

\usepackage{braket}

\usepackage{graphicx}
\usepackage{dcolumn}
\usepackage{bm}
\usepackage[dvipsnames]{xcolor}
\usepackage{svg}
\usepackage{float}
\usepackage{soul}
\usepackage{comment}
\usepackage{ulem}
\usepackage[english]{babel}
\begin{document}
\title{Active Leakage Cancellation in Single Qubit Gates}

\author{Ben Chiaro}
\email{equal contribution: chiaro@google.com}
\author{Yaxing Zhang}
\email{equal contribution: yaxingzhang@google.com}
\affiliation{Google Quantum AI, Santa Barbara, California 93111, USA}

\date{\today}
\begin{abstract}
The ability to perform fast and accurate rotations between the computational basis states of quantum bits is one of the most fundamental requirements for building a quantum computer. Because physical qubits generally contain more than two levels, faster gates often result in  a higher leakage rate outside of the computational space. In this letter, we enhance the state-of-the-art single qubit gate by introducing active leakage cancellation.  This is accomplished via a second drive tone near the leakage transition such that we cancel the leakage caused by the main drive.  Furthermore, we describe a measurement sequence that can be used to calibrate the parameters of this leakage cancellation drive. Finally, we apply the technique to superconducting transmon qubits and achieve leakage suppression by up to a factor of 20 when compared to pulses shaped with the standard DRAG technique. This results in a coherence-limited gate infidelity of $7.5\times 10^{-5}$ with leakage below $10^{-5}$ level, for a 10 ns $\pi/2$ gate and 196 MHz qubit anharmonicity.
\end{abstract}

\maketitle

\section{Introduction}
While below-threshold quantum error correction (QEC) performance has been demonstrated recently~\cite{google2023suppressing,aiquantum_below_threshold, putterman_hardware-efficient_2025}, further advancement requires improving the performance of the physical qubit quantum gates. Weakly anharmonic qubits such as superconducting transmons~\cite{koch2007charge,schreier2008suppressing, peterer2015coherence,kono_breaking_2020,place2021new,  lazuar2023calibration}, have similar transition frequencies for $\ket{0} \rightarrow \ket{1}$ and $\ket{1} \rightarrow \ket{2}$. These qubits face a fundamental tension in gate optimization:  It is desirable to make the operations as fast as possible to minimize the impact of decoherence, however, as gate length is reduced, power from the driving field is increasingly delivered to the undesired transition from $\ket{1}$ to $\ket{2}$.  This increases leakage outside the computational subspace which is detrimental to QEC~\cite{miao2023overcoming}.

There have been many attempts to engineer the control pulse spectrum near the frequency of the $\ket{1} \rightarrow \ket{2}$ transition $f_{21}$ to reduce leakage.  Of note is Derivative Removal by Adiabatic Gate  (DRAG), which produces a spectral notch at the specified frequency by adding a quadrature component to the control envelope \cite{motzoi_simple_2009,gambetta_analytic_2011, lucero2010reduced,chow2010optimized, DRAG_RB}. More recently, Fourier ansatz spectrum tuning (FAST) has been developed to further reduce power near $f_{21}$ with a control envelope that comprises a linear combination of higher harmonics of the base cosine envelope \cite{FAST}. While FAST has been shown to suppress leakage beyond the standard cosine with DRAG, the harmonics imply a more stringent requirement on the pulse generator's sampling rate.

In this work, we introduce a simple yet powerful control strategy to suppress leakage in single qubit gates.  Our strategy is to implement an {\it active leakage cancellation} (ALC) drive signal at the dominant leakage transition frequency $f_{21}$, simultaneous with the primary drive, that cancels the leakage caused by the main drive.  The technique is applicable to all qubit systems supporting states outside the computational basis, and especially relevant to low-anharmonicity systems.  We demonstrate the technique using superconducting transmon qubits and achieve 10 to 20 fold reduction in leakage per single qubit $\pi/2$ gate, compared to standard DRAG, across qubits with low (158 MHz) to moderate (196 MHz) anharmonicity. On a 196 MHz anharmonicity qubit, We push the leakage below $10^{-5}$ for a 10 ns $\pi/2$ gate, compared to $10^{-4}$ using only DRAG.

In contrast to FAST~\cite{FAST} and higher-derivative DRAG~\cite{DRAG_HD}, the power spectrum of the ALC pulse is well confined to the vicinity of the targeted transitions and does not increase away from them.  This feature implies less stringent requirements on the pulse generator.  We accomplish this by adding only two parameters to the pulse envelope, the cancellation drive amplitude and frequency, that can be calibrated independent of the primary drive. This allows for a simple calibration procedure compared to other pulse shaping techniques such as those based on optimal control~\cite{werninghaus2021leakage}.

\begin{figure}[t]
    \centering
    \includegraphics[width = 0.48\textwidth]{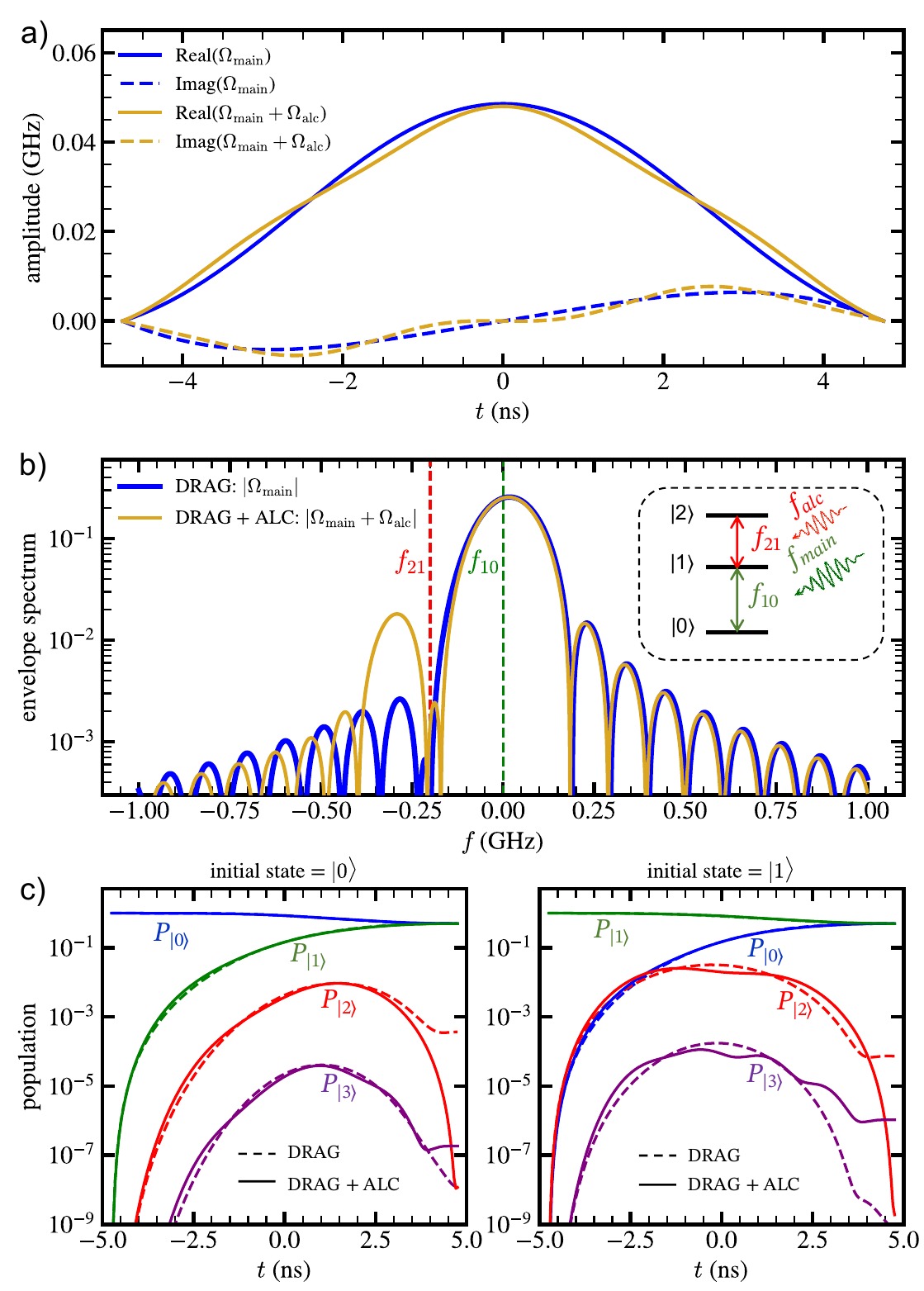}
    \caption{Pulse envelope of the ALC drive for a $\pi/2$ gate with $\eta = 200$ MHz and $t_{\rm gate} = 9.5$ ns.  a) Time-domain pulse envelope components. The real part of the primary drive (solid, blue) follows a raised-cosine pulse shape, and the imaginary part (dashed, blue) results from the application of DRAG. The composite pulse (primary + ALC drive) is shown in gold.  b) The spectra of the pulse envelope for the basic raised cosine with DRAG (blue), and with added ALC (gold) over a 2 GHz band centered about $f_{10}$. Vertical dashed lines indicate transition frequencies $f_{21}$ (red) and $f_{10}$ (green). Inset shows the level diagram of a transmon qubit with the primary and ALC drive: main drive frequency $f_{\rm main}$ is close to $f_{10}$ while the ALC drive frequency $f_{\rm alc}$ is close to $f_{21}$. c) qubit states $|0\rangle$ (blue), $|1\rangle$ (green) and $|2\rangle$ (red) and $|3\rangle$ (purple) populations during the gate operation using DRAG only (dashed) vs using DRAG and ALC (solid), for qubit initial state in $|0\rangle$ (left) and $|1\rangle$ (right). Final population in state $|2\rangle$ is substantially reduced by using DRAG + ALC versus DRAG only.}
    \label{fig:Concept}
\end{figure}

\section{Pulse shaping and modeling}

We describe ALC pulse shaping and modeling for transmon qubits, a specific case of this broadly applicable technique.
We model the transmon as a Kerr nonlinear oscillator~\cite{blais2021circuit}. Its Hamiltonian is $H_q/h = f_{10}a^\dagger a - \eta a^{\dagger 2}a^2/2$ where $a,a^\dagger$ are bosonic ladder operators, $f_{10}$ is the qubit transition frequency from state $\ket{0}$ to $\ket{1}$, $\eta$ is the anharmonicity that sets the difference between transition frequency from $\ket{2}$ to $\ket{1}$ and $\ket{1}$ to $\ket{0}$: $\eta = f_{10} - f_{21}$. For transmons, we have $\eta > 0$ and $\eta \ll f_{10}$.  
We describe the dynamics during microwave drive in a frame that rotates at $f_{10}$.  Under the rotating wave approximation, the Hamiltonian reads (see Appendix~\ref{sec:lab_frame} for the lab-frame Hamiltonian):
\begin{align}
\label{eq:hamiltonian}
H(t)/h = -\frac{\eta}{2}\hat{a}^{\dagger 2} \hat{a}^{2}& + \frac{1}{2}\left[\Omega_{\rm main}(t) + \Omega_{\rm alc}(t)\right] \hat{a} \nonumber \\
&+ \frac{1}{2}\left[\Omega_{\rm main}^{*}(t) + \Omega_{\rm alc}^*(t)\right] \hat{a}^\dagger, \end{align}
where $\Omega_{\rm main}(t)$ and $\Omega_{\rm alc}(t)$ refer to the complex envelope of the primary drive and ALC drive, respectively. In all equations, we use lower-case ``alc" to indicate active leakage cancellation. 

Our primary control envelope is the same as in Ref.~\cite{DRAG_RB}. As in the standard DRAG scheme, we take this envelope to be real valued in the rotating frame of the qubit and introduce a notch in the spectrum at $f_{21}$ by adding a quadrature component proportional to the derivative of the in-phase component:
\begin{align}
\label{eq:main_drive}
    \Omega_{\rm main}(t) = A_{\rm main}\left[F(t) +i\frac{\alpha}{2\pi\Delta}\frac{dF(t)}{dt} \right] e^{i2\pi\delta_{\rm main} t},
\end{align}
where $A_{\rm main}$ and $\delta_{\rm main}$ are the primary drive amplitude and detuning from $f_{10}$, i.e., $\delta_{\rm main} = f_{\rm main}-f_{10}$. This detuning compensates the drive-induced ac Stark shift~\cite{DRAG_RB}.
In this work, we fix the DRAG parameter $\alpha$ to be 1 to minimize leakage. The parameter $\Delta$ sets the notch position and is fixed to be $\Delta= -\eta - \delta_{\rm main}$. We consider the common choice of a raised cosine envelope for single qubit gates,
\begin{align}
\label{eq:raised_cosine}
    F(t) =1 - \cos\left[ 2 \pi \frac{t+t_{\rm gate}/2} {t_{\rm gate}}\right],
\end{align} 
which starts at $t=-t_{\rm gate}/2$ and ends at $t=t_{\rm gate}/2$. We choose a full raised cosine because both the pulse and its derivative start and end at zero and are continuous everywhere, which is convenient for numerical and analytical study.

Whereas in DRAG, we suppress the spectrum of the drive pulse near $f_{21}$ by adding a quadrature component at the same frequency as the primary drive, ALC introduces an additional drive tone with frequency near $f_{21}$.  The envelope of this additional drive is: 
\begin{align}
\Omega_{\rm alc}(t) = iA_{\rm alc} \frac{1}{2\pi \delta_{\rm alc}} \frac{d F(t)}{dt}e^{i 2\pi \delta_{\rm alc}t + i\phi_{\rm alc}}.
\label{eq:AL_drive}
\end{align}

The amplitude $A_{\rm alc}$ and detuning $\delta_{\rm alc} = f_{\rm alc}-f_{10}$ of the ALC drive are optimized to minimize leakage (Sec.~\ref{sec:calibration}), while the relative phase between the main and cancellation drives $\phi_{\rm alc}$ is held fixed at zero (Appendix~\ref{app:analytics}). Typically we have $\delta_{\rm alc} \approx -\eta$, and amplitude $A_{\rm alc}$ is $\sim 10\%$ to 20$\%$ of the main pulse amplitude $A_{\rm main}$ with opposite sign, for gates between $\sim 10$ and 15 ns. 

The ALC drive is similar to the DRAG term of the main drive, but is shifted in frequency by approximately $-\eta$.  It has a notch near $f_{21}$, as shown in Eq.~(\ref{eq:spectrum_alc}), which, when combined with DRAG, strongly suppresses leakage.

Figure~\ref{fig:Concept} illustrates the construction of the leakage cancelled gate for $F(t)$ given by Eq.~(\ref{eq:raised_cosine}).  Pulse parameters are numerically optimized to minimize $\pi/2$ gate error. In Fig.~\ref{fig:Concept}(a), shown in blue is the time domain control envelope of the main drive optimized without the ALC drive. The real part of this envelope (solid) is due to the raised cosine profile while the imaginary part (dashed) results from the application of DRAG.  The gold lines in Fig.~\ref{fig:Concept}(a) show the composite pulse envelope including the ALC drive. We numerically verified that adding the ALC drive only weakly changes the optimal main pulse parameters. Furthermore, the strength of the ALC drive is weak compared to the main drive, as indicated by the small difference between main and the composite drive. Unlike the main drive envelope, the composite envelope is not a simple sine wave.  This is because the additional ALC drive has modulation at frequency $\delta_{\rm alc}$ on top of the raised cosine envelope. 

Figure~\ref{fig:Concept}(b) shows the spectra of the pulse envelopes, computed as the Fourier transform of the time-domain pulse. It follows from Eqs.~(\ref{eq:main_drive}, \ref{eq:AL_drive}) that the spectra of the primary and ALC drive are: 
\begin{align}
    \Omega_{\rm main}[f]&= A_{\rm main} \left[ 1 - \frac{\alpha}{\Delta}(f - \delta_{\rm main})\right]F[f-\delta_{\rm main}],\\
    \Omega_{\rm alc}[f]&= A_{\rm alc} \left[ 1 - \frac{1}{\delta_{\rm alc}}f\right]F[f-\delta_{\rm alc}],
    \label{eq:spectrum_alc}
\end{align}
where $\Omega_{\rm main}[f] \equiv \int_{-\infty}^\infty \Omega_{\rm main}(t)e^{-i2\pi f t} dt $, $\Omega_{\rm alc}[f] \equiv \int_{-\infty}^\infty \Omega_{\rm alc}(t)e^{-i2\pi f t} dt $, and $F[f] \equiv \int_{-\infty}^\infty F(t)e^{-i2\pi f t} dt $.
We fix $\alpha = 1$, and $\Delta = -(\eta + \delta_{\rm main})$ such that spectrum $\Omega_{\rm main}[f]$ has a spectral notch at $f = -\eta$. The ALC drive itself has a notch at $\delta_{\rm alc}$ which  is near $-\eta$. Compared to the spectrum of the main drive (DRAG only, blue), the composite pulse (DRAG + ALC, gold) has a larger spectral weight few hundreds of MHz below the leakage transition $f_{21}$. However, farther away from $f_{21}$, in particular, near the intended transition $f_{10}$, the spectrum is only weakly perturbed compared to the spectrum without ALC drive. 

Figure~\ref{fig:Concept}(c) shows numerically simulated qubit state populations during the gate, for qubit initially in state $|0\rangle$ and $|1\rangle$, respectively. While state $|2\rangle$ population {\it during} the gate is similar between pulses with only DRAG and with DRAG plus ALC, the final population in state $|2\rangle$ is reduced by several orders of magnitude by applying the ALC drive.  The final $|3\rangle$ state population is higher in DRAG + ALC, but remains orders of magnitude smaller than state $|2\rangle$ population in the case of DRAG only.

In Appendix~\ref{app:analytics}, we show analytically that the ALC drive cancels leakage through destructive interference between the state $|2\rangle$ amplitude caused by the main drive and by the ALC drive. This is achieved through tuning the ALC drive amplitude and frequency. Analytical expressions for the optimal ALC drive amplitude and frequency agree with numerical results.

\section{Calibration}
\label{sec:calibration}
\begin{figure}[t]
    \centering
    \includegraphics{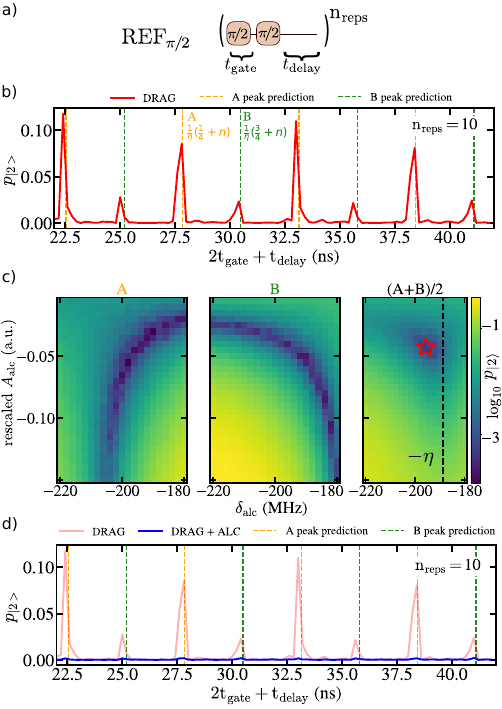}
    \caption{Calibration sequence for ALC drive parameters.  a) The pulse sequence for the $\rm REF_{\pi/2}$. b) Example data from the $\rm REF_{\pi/2}$ for the case of $n_{\rm reps}=10$, the peaks are due to the amplification of coherent leakage induced by the $\pi/2$ pulses.  c) Optimization of ALC drive amplitude $A_{\rm alc}$ and detuning $\delta_{\rm alc}$. The left and center panels are acquired at two different $t_{\rm delay}$ corresponding to a type-A and type-B peak from b), respectively.  The right panel shows the average leakage population $p_{|2>}$. The optimal ALC parameters (red star) are where the average of leakage contributions from A and B peaks are minimized. The optimal $\delta_{\rm al}$ is slightly shifted from $-\eta$ (vertical dashed line). d) Uncompensated data (red) from the $\rm REF_{\pi/2}$ from panel (b) (red) compared with the $\rm REF_{\pi/2}$ with ALC (blue) showing suppression of coherent leakage.}
    \label{fig:calibrate}
\end{figure}

In this section, we detail an experimental procedure to determine the optimal ALC parameters.   
Our calibration sequence employs the Ramsey error filter (REF)~\cite{REF_Lucero} technique to amplify the coherent leakage produced by the control pulse.  This leakage is minimized for optimal ALC parameters.  The pulse sequence for the $\pi/2$ variant of the Ramsey error filter REF$_{\pi/2}$ is shown in Fig.~\ref{fig:calibrate}(a).  The $\pi/2$ pulses drive a small population from $\ket{1}$ to $\ket{2}$ while the Ramsey delay $t_{\rm delay}$ generates phase accumulation in the $\ket{1}$, $\ket{2}$ subspace; the delay can also be implemented using virtual Z rotations. This phase accumulation allows the constructive interference between the pulse unitaries and thus population accrues into the $\ket{2}$ state.  This sequence is repeated $n_{\rm reps}$ times to amplify the population transfer.  

In Fig.~\ref{fig:calibrate}(b), we show the observed $\ket{2}$ population $p_{\ket{2}}$ vs $t_{\rm delay}$ during the REF$_{\pi/2}$ experiment.  Peaks in $p_{\ket{2}}$ correspond to the constructive interference condition: $\eta(4t_{\rm gate}+2t_{\rm delay})= (k+1/2)$, with $k=0,1,2..$. This condition can be understood as follows. During every four $\pi/2$ pulses (consider this as a cycle), the qubit goes from $\ket{0}$ to $\ket{1}$ and returns to $\ket{0}$ with a $\pi$ phase shift. To have constructive interference in state $\ket{2}$ amplitude from cycle to cycle, we need to have $\eta$ multiplied by the cycle time $(4t_{\rm gate}+2t_{\rm delay})$ to equal half-odd-integers. The peaks corresponding to $k$ being even ($k=2n$) or odd integer ($k=2n+1$) can in general be different, they are referred to as type-A and type-B peaks, respectively, in Fig.~\ref{fig:calibrate}(b). Their heights are proportional to the leakage probability accumulated in state $|2\rangle$ within a cycle (four $\pi/2$ pulses) and approximately scale as $n_{\rm reps}^2$~\cite{sm}.

Figure~\ref{fig:calibrate}(c) shows the two-parameter optimization of the ALC drive parameters. In the left subpanel, we show the observed $p_{\ket{2}}$ with $t_{\rm delay}$ held fixed at the value that maximizes $p_{\ket{2}}$ for a type-A peak.  The center subpanel shows the data for a type-B peak.  A naive optimization might minimize the population in either type-A peaks or type-B peaks.  However this generally fails because the targeted peak may be minimized while peaks of the other type are enhanced.  In the right-most panel of Fig~\ref{fig:calibrate}(c), we show the average $p_{\ket{2}}$ in the A and B peaks vs $A_{\rm alc}$ and $\delta_{\rm alc}$. The red star indicates the optimal ALC drive parameters. As expected, the optimal detuning is near $-\eta$. In Fig.~\ref{fig:calibrate}(d), we show validation data (blue) demonstrating that these cancellation parameters suppress $\pi/2$ gate leakage for all $t_{\rm delay}$.

\section{Demonstration}
\label{sec:demonstration}
\begin{figure}[t]
    \centering
    \includegraphics[width=0.48\textwidth]{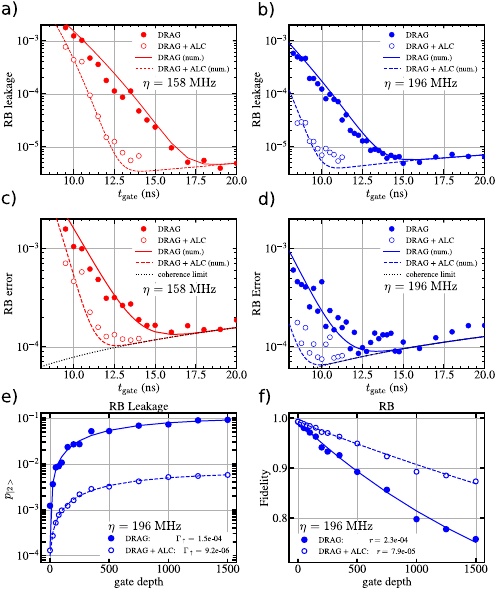}
    \caption{Randomized benchmarking of single qubit $\pi/2$ gates with ALC, for qubits with $\eta = $ 158\,MHz ($f_{10} = 4.5$\,GHz) and 196\,MHz ($f_{10} = 5.7$\,GHz). a) and b) The leakage rate for gates constructed with DRAG only (solid markers) and with both DRAG and ALC (hollow markers) as observed through RB sequences. Measured leakage rates agree with numerical predictions (lines) of coherent leakage with a background rate from incoherent heating. c) and d) Average gate error as assessed by RB. The numerical prediction (lines) reflects the gate error through the leakage channel plus error from qubit decoherence due to $T_1,T_{\phi}$ (black dotted line). The latter uses formula, gate error = $t_{\rm gate}/ 3T_1 + t_{\rm gate}/3 T_{\phi}$ with independently measured $T_1$ and $T_{\phi}$ [panel (c): $T_1$ = 104 ${\rm \mu s}$, $T_{\phi}$ = 118 ${\rm \mu s}$; panel (d) $T_1$ = 101 $\mu s$,$T_{\phi}$ = 109 ${\rm \mu s}$] In panel (c), the result from the formula is multiplied by a factor of 1.3 to match the data indicating additional decoherence during the gate compared to idle. In panel (d), the formula agrees with the measured error at large $t_{\rm gate}$ without correction. e) and f) state $\ket{2}$ population and RB fidelity versus gate depth for $t_{\rm gate} = 9.75$ ns and $\eta = 196$~MHz, with legends indicating leakage per cycle $\Gamma_{\uparrow}$ and RB error $r$.}
    \label{fig:demonstrate}
\end{figure}

We demonstrate our technique on qubits with 158~MHz and 196~MHz nonlinearity, using an upconverted 1~GS/s AWG.  Figure~\ref{fig:demonstrate} shows the performance of $\pi/2$ gates vs $t_{\rm gate}$ with and without ALC. For each $t_{\rm gate}$, we calibrate the primary drive parameters without ALC, including $A_{\rm main}$ and $\delta_{\rm main}$ and a post-gate phase correction. Once the primary drive has been calibrated, we calibrate $A_{\rm alc}$ and $\delta_{\rm alc}$ using the procedure detailed in section \ref{sec:calibration}.  Finally, we re-calibrate the primary drive parameters with ALC. The main drive calibration sequence takes three minutes, as does the ALC optimization. Replacing the 2D optimization in Fig.\ref{fig:calibrate}(c) with a model-guided approach may reduce this time.

Figure~\ref{fig:demonstrate}(a,b) compare the leakage probability per gate with and without ALC, during randomized benchmarking (RB). While the leakage rate generally increases with the decrease in the gate time, without ALC (solid markers), the leakage rate increases beyond the $10^{-5}$ level approximately at $t_{\rm gate}\approx 2.5/\eta$, this corresponds to $t_{\rm gate}\approx 16$~ns for $\eta = 158$~MHz, and 13 ns for $\eta = 196$~MHz. With ALC (hollow markers), we push this to a significantly smaller gate time of $t_{\rm gate}\approx 1.9/\eta$, approximately $t_{\rm gate}\approx 12$~ns for $\eta = 158$~MHz, and 9.7 ns for $\eta = 196$~MHz. For fixed $t_{\rm gate}$, ALC reduces the leakage rate by up to 20x compared to gates without ALC.

At longer gate times, the leakage rate is limited by incoherent heating causing a linear rise of leakage with $t_{\rm gate}$.  On the $\eta = 196$~MHz device,  independently measured steady-state level populations imply an effective temperature of $84.2$~mK.  This agrees with the $82.0$~mK temperature inferred from the leakage per gate during RB. We extract an effective temperature of $57.7$~mK from the leakage rates on the $\eta = 158$~MHz device. At these larger $t_{\rm gate}$, coherent leakage is too small to be detected by our REF protocol which prevents ALC calibration, therefore we do not report ALC leakage rates above certain $t_{\rm gate}$ in Fig.~\ref{fig:demonstrate}.

Figure~\ref{fig:demonstrate}(c,d) compare the average gate error with and without ALC. At large $t_{\rm gate}$, the gate error is limited by decoherence (black dotted lines) and scales approximately linearly with $t_{\rm gate}$. At smaller $t_{\rm gate}$, error increases in response to increasing coherent leakage.   In this regime, we observe a significant error rate reduction with ALC. For $\eta = 158$~MHz, we achieve a gate error of $1.3\times 10^{-4}$ at $t_{\rm gate} = 12$~ns, compared to $3.1\times 10^{-4}$ at the same gate time using DRAG alone, a factor of 2.4 reduction. We attribute this improvement to the 12x reduction in leakage, from $1.8\times 10^{-4}$ to $1.5\times 10^{-5}$, due to applying ALC.  Similarly, for $\eta = 196$~MHz with $t_{\rm gate} = 9.75$~ns, we observe a 16x leakage rate reduction from $1.5\times 10^{-4}$ without ALC to $9.2\times 10^{-6}$ with ALC with a concomitant improvement in the gate error from $2.3\times 10^{-4}$ without ALC to $7.9\times 10^{-5}$ with ALC, a factor of 2.9 reduction. With ALC, the lowest gate error we observe is $7.5\times 10^{-5}$ at $t_{\rm gate} = 10$~ns, which is primarily coherence limited. For both values of $\eta$, ALC permits a higher fidelity gate than would be possible with DRAG alone.

Figure~\ref{fig:demonstrate}(e,f) highlights the leakage population and RB fidelity versus gate depth at $t_{\rm gate} = 9.75$~ns for $\eta = 196$~MHz. The gate fidelity and leakage rate improvements from ALC drive persist at large gate depths. 

Numerical results in Fig.~\ref{fig:demonstrate} are obtained by simulating the rotating-frame Hamiltonian in Eq.~(\ref{eq:hamiltonian}). It follows the iterative experimental procedure of section \ref{sec:calibration} to optimize the primary and ALC drive parameters independently. The simulated leakage rate agrees with experiments as shown in Fig.~\ref{fig:demonstrate}(a,b). Simulations show that globally optimizing the primary and ALC drive parameters could achieve another two to eight-fold reduction in the leakage rate~\cite{sm}, suggesting calibration procedure refinement could further reduce experimental leakage rates. 

\section{Conclusions}
We developed an active leakage cancellation strategy that significantly enhances the performance of single-qubit gates, particularly in systems with low anharmonicity. By implementing a second drive tone at the leakage transition frequency $f_{21}$, we efficiently compensate the leakage caused by the primary drive. We developed a simple calibration procedure to optimize the leakage cancellation pulse parameters. Moreover, we experimentally demonstrated this technique on superconducting transmon qubits, and achieved a 10 to 20-fold reduction in leakage compared to standard DRAG. For qubits with an anharmonicity of 196 MHz, leakage is suppressed below $10^{-5}$ for a 10 ns $\pi/2$ gate, enabling a coherence-limited gate infidelity of $7.5 \times 10^{-5}$. Through further optimization of the leakage cancellation pulse shape, more sophisticated calibration procedure, or combining the technique with other pulse shaping technique such as FAST, we anticipate even stronger performance. This active leakage cancellation strategy represents a simple yet powerful approach to improving the speed and accuracy of single-qubit gates, much needed for the advancements towards practical quantum computing.

\begin{acknowledgments}
We are grateful to the Google Quantum AI team for building, operating, and maintaining the software and hardware infrastructure used in this work. We thank Alexander N. Korotkov and Kenny Lee for the support and valuable discussions throughout the project.

\end{acknowledgments}

\begin{appendix}
\section{Hamiltonian in the lab frame}
\label{sec:lab_frame}
In this section, we describe the Hamiltonian of the driven transmon in the lab frame. 

In the presence of both primary and ALC drive, the transmon Hamiltonian in the lab frame is given by
\begin{align}
    H_{\rm lab}(t)/h&= f_{10}a^\dagger a - \eta a^{\dagger 2} a^2/2 \nonumber \\&+ \Big[A_{\rm main}F(t) \cos(2\pi f_{\rm main}t) \nonumber \\&-A_{\rm main}\frac{\alpha}{2\pi\Delta} \dot F(t)  \sin(2\pi f_{\rm main}t)\nonumber \\
    &- A_{\rm alc}\frac{1}{2\pi\delta_{\rm alc}} \dot F(t)\sin(2\pi f_{\rm alc}t)\Big] (a + a^\dagger),
\label{eq:H_lab}
\end{align}
where $\delta_{\rm alc} = f_{\rm alc} - f_{10}$. The first two terms in the square bracket describe the primary drive and its DRAG component and the third term describes the ALC drive. Frequencies of the main drive and ALC drive are $f_{\rm main}$ and $f_{\rm alc}$, respectively. Function $F(t)$ describes the pulse envelope. In the frame that rotates at qubit frequency $f_{10}$ and under rotating wave approximation, this Hamiltonian transforms to the Hamiltonian in Eq.~(\ref{eq:hamiltonian}) in which $\Omega_{\rm main}$ and $\Omega_{\rm alc}$ are given by Eq.~(\ref{eq:main_drive}) and Eq.~(\ref{eq:AL_drive}), respectively.

\section{Analytical analysis of the ALC drive}
\label{app:analytics}

\begin{figure}[t]
    \centering
    \includegraphics[width = 0.24\textwidth]{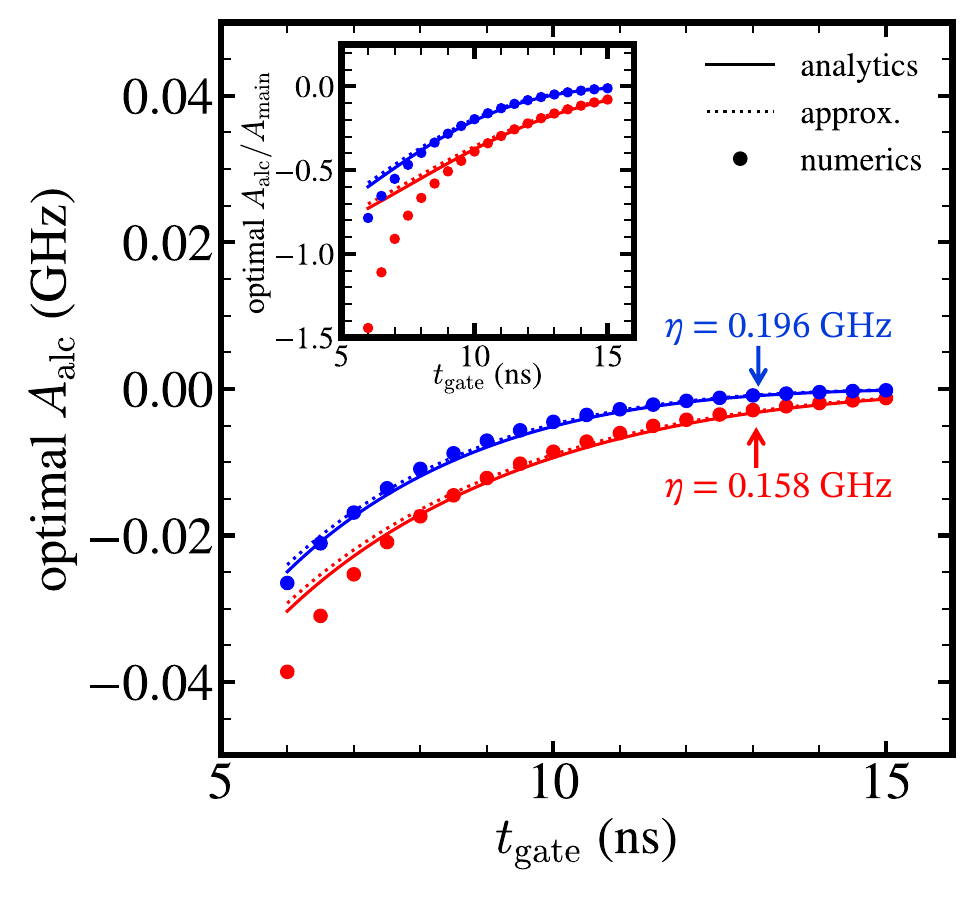}\hfill
    \includegraphics[width = 0.24\textwidth]{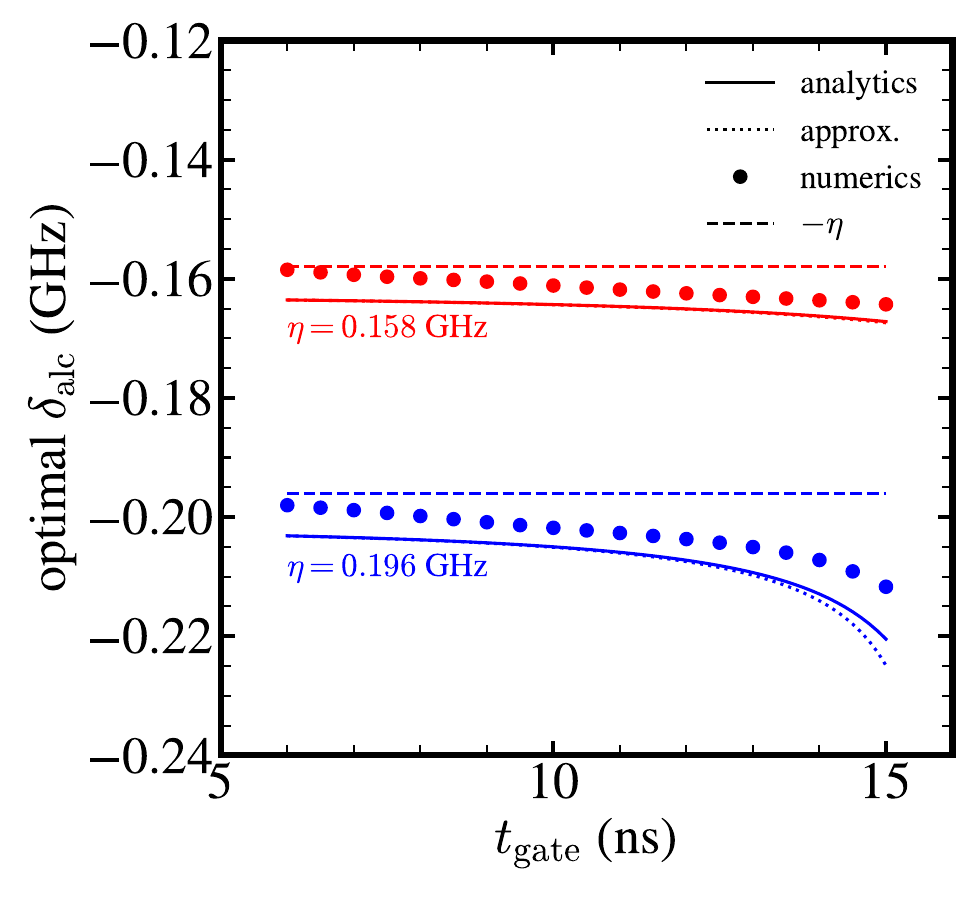}
    \caption{Optimal ALC drive amplitude (left) and detuning (right) for various gate times. Solid lines show analytical results found through solving Eqs.~(\ref{eq:condition_1_final},\ref{eq:condition_2_final}); dotted lines show results using Eqs.~(\ref{eq:expression_A_alc},\ref{eq:expression_delta_alc}) which provide approximate solutions to Eqs.~(\ref{eq:condition_1_final},\ref{eq:condition_2_final}); markers show numerical simulation results obtained through an iterative optimization procedure similar to the experimental calibration procedure described in the beginning of Sec.~\ref{sec:demonstration}. Different colors refer to different anharmonicities: $\eta=0.158$~GHz (red), $\eta = 0.196$~GHz (blue). Inset of left panle shows the ratio of the ALC drive amplitude over the main drive amplitude. Dashed lines in the right panel indicate the location of $-\eta$.}
    \label{fig:optimal_parameters}
\end{figure}

In this section, we provide analytical results for the optimal parameters of the ALC drive, and show explicitly how the cancellation drive works through destructive interference of the leakage amplitude. Analysis in this section applies to general pulse shape $F(t)$ as long as the pulse is symmetric with respect to the pulse center.  

Since the leakage is dominated by leakage to state $|2\rangle$, for analytical analysis, we disregard states higher than $|2\rangle$, and write the time-dependent wave function as $|\psi(t)\rangle = \sum_{n=0,1,2}c_n(t)|n\rangle$. It follows from the Schr\"odinger equation and Hamiltonian in Eq.~(\ref{eq:hamiltonian}) that state $|2\rangle$ amplitude $c_2(t)$ at the end of the gate is given by:
\begin{align}
    c_2\left(\frac{t_{\rm gate}}{2}\right) = & -i \sqrt{2}\pi\int_{-\frac{t_{\rm gate}}{2}}^{\frac{t_{\rm gate}}{2}} e^{i2\pi\eta(\frac{t_{\rm gate}}{2}-t)}\nonumber \\
   & \times [\Omega^*_{\rm main}(t) + \Omega^{*}_{\rm alc}(t)] c_1(t)dt \nonumber \\
     = -i\frac{\sqrt{2}}{2}\int_{-\frac{t_{\rm gate}}{2}}^{\frac{t_{\rm gate}}{2}} &e^{i2\pi\eta(\frac{t_{\rm gate}}{2}-t)}\Bigg[-i\frac{A_{\rm main}}{\eta}F(t)\dot c_1(t) \nonumber\\- i\frac{A_{\rm alc}}{\delta_{\rm alc}}\dot F(t)&e^{-i 2\pi \delta_{\rm alc}t-i\phi_{\rm alc}}c_1(t)\Bigg]dt
\label{eq:c2}
\end{align}
In the second line, we have substituted the expressions for $\Omega_{\rm main}$ and $\Omega_{\rm alc}$ in Eqs.~(\ref{eq:main_drive},\ref{eq:AL_drive}) and performed integration by part for the main drive. For simplicity, we have disregarded the small detuning $\delta_{\rm main}$ from the main drive as well as the ac Stark shift of state $|2\rangle$ due to the off-resonant drive acting on state $|2\rangle$ to $|3\rangle$ transtion, their magnitudes are typically much smaller than the anharmonicity $\eta$ itself. 
Equation~(\ref{eq:c2}) shows that the leakage amplitude to state $|2\rangle$ is not simply related to the Fourier transform of the drive envelope at the leakage transition frequency, but also depends on the time-dependent amplitude in state $|1\rangle$. 

To simplify Eq.~(\ref{eq:c2}), we make the approximation that state $|1\rangle$ amplitude $c_1(t)$ follows the ideal evolution: 
\begin{align}
\label{eq:c1}
    c_1(t) &\approx c_1\left(-\frac{t_{\rm gate}}{2}\right)\cos\left[\frac{\theta(t)}{2}\right]\nonumber\\ &-ic_0\left(-\frac{t_{\rm gate}}{2}\right)\sin\left[\frac{\theta(t)}{2}\right],\\
\label{eq:theta_t}
    \theta(t)& \equiv \frac{\pi}{2t_{\rm gate}}\int_{-t_{\rm gate}/2}^t F(t')dt' =\pi/4 + \tilde \theta(t), 
    \\
    \tilde\theta(t) & \equiv \frac{\pi}{2t_{\rm gate}}\int_0^t F(t')dt'.
\end{align}
Here $\theta(t)$ is the time-dependent rotation angle of the Bloch vector, it increases from 0 to $\pi/2$ from the beginning to the end of the $\pi/2$ gate. In Eq.~(\ref{eq:theta_t}), we express $\theta(t)$ in terms of $\tilde \theta(t)$ because for symmetric pulse, $\tilde \theta(t)$ is antisymmetric with respect to $t=0$ [i.e., $\tilde \theta(-t) = -\tilde \theta(t)$], which simplifies the subsequent calculation. Without loss of generality, we have chosen the normalization condition that the area of $F(t)$ under the pulse is equal to one.  

To cancel leakage for arbitrary initial state in the computational subspace, we must have $c_2(t_{\rm gate}/2) = 0$ for two orthogonal initial states. It is convenient to choose them to be $(|1\rangle \pm |0\rangle)/\sqrt{2}$. For these two initial states, state $|1\rangle$ amplitude $c_1(t)$ is given by $\exp[\mp i\theta(t)/2]$. Substituting these $c_1(t)$ to Eq.~(\ref{eq:c2}) and requiring $c_2(t_{\rm gate}/2) = 0$ give us the following two leakage cancellation conditions that need to be satisfied simultaneously: 
\begin{widetext}
\begin{align}
\label{eq:condition_1}
  -\frac{\pi}{4t_{\rm gate}} \frac{A_{\rm main}}{\eta} \int_{-\frac{t_{\rm gate}}{2}}^{\frac{t_{\rm gate}}{2}} e^{-i2\pi\eta t}F^2(t) e^{-i\tilde\theta(t)/2}dt &= i\frac{A_{\rm alc}}{\delta_{\rm alc}} \int_{-\frac{t_{\rm gate}}{2}}^{\frac{t_{\rm gate}}{2}} e^{-i2\pi(\eta+\delta_{\rm alc})t-i\phi_{\rm alc}}\dot F(t) e^{-i\tilde\theta(t)/2}dt  \\
\frac{\pi}{4t_{\rm gate}} \frac{A_{\rm main}}{\eta} \int_{-\frac{t_{\rm gate}}{2}}^{\frac{t_{\rm gate}}{2}} e^{-i2\pi\eta t}F^2(t) e^{i\tilde\theta(t)/2}dt &= i\frac{A_{\rm alc}}{\delta_{\rm alc}} \int_{-\frac{t_{\rm gate}}{2}}^{\frac{t_{\rm gate}}{2}} e^{-i2\pi(\eta+\delta_{\rm alc})t-i\phi_{\rm alc}}\dot F(t) e^{i\tilde\theta(t)/2}dt 
\label{eq:condition_2}
\end{align}
At first glance, it appears that we need to tune four parameters to satisfy the above two complex equations simultaneously. It turns out that symmetry of the pulse with respect to the pulse center reduces the number of parameters needed to three (phase, detuning and amplitude of the ALC drive). First of all, due to the symmetry, left-hand sides of Eqs.~(\ref{eq:condition_1},\ref{eq:condition_2}) are purely real while the right-hand sides excluding the phase factor $\exp(-i\phi_{\rm alc})$ are also real. This means that to satisfy the above equations, phase $\phi_{\rm alc}$ has to be zero. After setting $\phi_{\rm alc}$ to zero, the above equations can be rewritten as 
\begin{align}
      -\frac{\pi}{4t_{\rm gate}} \frac{A_{\rm main}}{\eta} \int_{0}^{\frac{t_{\rm gate}}{2}} F^2(t) \cos\left[\tilde\theta(t)/2+2\pi \eta t\right]dt &= \frac{A_{\rm alc}}{\delta_{\rm alc}} \int_{0}^{\frac{t_{\rm gate}}{2}} \dot F(t) \sin\left[\tilde\theta(t)/2+2\pi(\eta+\delta_{\rm alc})t\right]dt  \\
      \frac{\pi}{4t_{\rm gate}} \frac{A_{\rm main}}{\eta} \int_{0}^{\frac{t_{\rm gate}}{2}} F^2(t) \cos\left[\tilde\theta(t)/2-2\pi \eta t\right]dt &= -\frac{A_{\rm alc}}{\delta_{\rm alc}} \int_{0}^{\frac{t_{\rm gate}}{2}} \dot F(t) \sin\left[\tilde\theta(t)/2-2\pi(\eta+\delta_{\rm alc})t\right]dt
\end{align}
Taking the sum and difference of the two equations above, we obtain the following:
\begin{align}
\label{eq:condition_1_final}
          &-\frac{\pi}{4t_{\rm gate}} \frac{A_{\rm main}}{\eta} \int_{0}^{\frac{t_{\rm gate}}{2}} F^2(t) \cos\left[\tilde\theta(t)/2\right]\cos\left(2\pi \eta t\right)dt = \frac{A_{\rm alc}}{\delta_{\rm alc}} \int_{0}^{\frac{t_{\rm gate}}{2}} \dot F(t) \sin\left[\tilde\theta(t)/2\right]\cos\left[2\pi(\eta+\delta_{\rm alc})t\right]dt \\
        &\frac{\pi}{4t_{\rm gate}} \frac{A_{\rm main}}{\eta} \int_{0}^{\frac{t_{\rm gate}}{2}} F^2(t) \sin\left[\tilde\theta(t)/2\right]\sin\left(2\pi\eta t\right)dt = \frac{A_{\rm alc}}{\delta_{\rm alc}} \int_{0}^{\frac{t_{\rm gate}}{2}} \dot F(t) \cos\left[\tilde\theta(t)/2\right]\sin\left[2\pi(\eta+\delta_{\rm alc})t\right]dt
\label{eq:condition_2_final}
\end{align}
Amplitude $A_{\rm main}$ of the main pulse can be approximated as $1/4t_{\rm gate}$ for $\pi/2$ pulse. The only unknowns are amplitude and detuning of the ALC drive. Therefore, given a pulse shape $F(t)$, Eqs.~(\ref{eq:condition_1_final},\ref{eq:condition_2_final}) allow us to solve for the optimal ALC drive parameters $A_{\rm alc}$ and $\delta_{\rm alc}$. 
These two equations are the main results of this section. 


Equations~(\ref{eq:condition_1_final},\ref{eq:condition_2_final}) can be reduced to explicit expressions for $\delta_{\rm alc}$ and $A_{\rm alc}$ by using the approximation that $\delta_{\rm alc}\approx -\eta$. Under this approximation, Eq.~(\ref{eq:condition_1_final}) becomes
\begin{align}
\label{eq:expression_A_alc}
    A_{\rm alc} \approx \frac{\pi}{(4t_{\rm gate})^2}\frac{\int_{0}^{\frac{t_{\rm gate}}{2}} F^2(t) \cos\left[\tilde\theta(t)/2\right]\cos\left(2\pi \eta t\right)dt} {\int_{0}^{\frac{t_{\rm gate}}{2}} \dot F(t) \sin\left[\tilde\theta(t)/2\right]dt}.
\end{align}
Dividing two sides of Eqs.~(\ref{eq:condition_1_final}, \ref{eq:condition_2_final}) by each other and keeping only linear term in $\delta_{\rm alc}+\eta$, we obtain that
\begin{align}
\label{eq:expression_delta_alc}
    \delta_{\rm alc} \approx -\eta -\frac{\int_{0}^{\frac{t_{\rm gate}}{2}} F^2(t) \sin\left[\tilde\theta(t)/2\right]\sin\left(2\pi\eta t\right)dt \int_{0}^{\frac{t_{\rm gate}}{2}} \dot F(t) \sin\left[\tilde\theta(t)/2\right]dt}{\int_{0}^{\frac{t_{\rm gate}}{2}} F^2(t) \cos\left[\tilde\theta(t)/2\right]\cos\left(2\pi \eta t\right)dt \int_{0}^{\frac{t_{\rm gate}}{2}} \dot F(t)(2\pi t) \cos\left[\tilde\theta(t)/2\right]dt}.
\end{align}
Because $\tilde \theta(t)$ has a maximum of $\pi/4$, numerator in the second term of the equation above is generally much smaller than the denominator, indicating that the initial assumption of $\delta_{\rm alc}\approx -\eta$ is valid. Equations~(\ref{eq:expression_A_alc},\ref{eq:expression_delta_alc}) provide close-form expressions for $A_{\rm alc}$ and $\delta_{\rm alc}$, respectively. We checked that they are very good approximation to Eqs.~(\ref{eq:condition_1_final},\ref{eq:condition_2_final}). 

\end{widetext}

Figure~\ref{fig:optimal_parameters} shows the comparison between optimal ALC drive parameters found through numerical simulations and analytical results, for two different qubit anharmonicities and for the pulse shape given by Eq.~(\ref{eq:raised_cosine}). We observe very good agreement between numerical and analytical results. The deviation in the drive amplitude at small gate times is likely due to that in the analytical derivation, we have neglected ac Stark shifts of the transmon levels due to the drive which become relatively large at short gate times.

\end{appendix}
\bibliographystyle{apsrev4-2}
%

\pagebreak
\clearpage

\setcounter{equation}{0}
\setcounter{section}{0}
\setcounter{figure}{0}
\setcounter{table}{0}
\setcounter{page}{1}
\makeatletter
\renewcommand{\theequation}{S\arabic{equation}}
\renewcommand{\thefigure}{S\arabic{figure}}
\renewcommand{\bibnumfmt}[1]{[S#1]}
\renewcommand{\citenumfont}[1]{S#1}
\onecolumngrid
\begin{center}
{\Large \textbf{Supplemental Material for the manuscript \\``Active Leakage Cancellation in Single Qubit Gates"} }
\end{center}
\vspace{2mm}
\twocolumngrid

\section{Explain interference peaks in Ramsey error filter}
\label{app:explain_AB_peaks}

In this section, we briefly explain the origin of the interference peaks in Fig.~2(b) of the main text and the interference condition. 

Let us consider the pulse sequence shown in Fig.~2(a) of the main text. During the first two $\pi/2$ pulses, qubit goes from state $|0\rangle$ to $|1\rangle$, and accumulates a small amplitude $c$ in state $|2\rangle$ due to coherent leakage. During the second two $\pi/2$ pulses, qubit goes from state $|1\rangle$ to $|0\rangle$, and accumulates additional amplitude $c'$ in state $|2\rangle$, which in general is different from $c$. At the end of the second repetition ($n_{\rm reps}=2$), the amplitude of state $|2\rangle$ can be written as follows:
\begin{align}
\label{eq:c2_nreps=2}
    c_2(n_{\rm reps}=2) = c e^{i2\pi \eta (2t_{\rm gate}+t_{\rm delay}) } + c'.
\end{align}
Because the two sets of $\pi/2$ pulses are separated by $2t_{\rm gate} + t_{\rm delay}$, there is an additional phase accumulated in front of the $c$ term due to the idle evolution. There is a common phase shift in front of $c$ and $c'$ term during $t_{\rm delay}$ after the second two $\pi/2$ pulses. This only causes an unimportant global phase shift in $c_2(n_{\rm reps}=2)$, which we have neglected. 

After each two repetitions (four $\pi/2$ pulses), state $|2\rangle$ amplitude has an additional $c_2(n_{\rm reps}=2)$ added to it, while existing state $|2\rangle$ amplitude gets an additional phase $2\pi \eta (4t_{\rm gate}+2t_{\rm delay})$ from the idle evolution. For each added $c_2(n_{\rm reps} = 2)$, there will be additional factor of $-1$ because every two repetition (i.e., four $\pi/2$ pulses), qubit completes a full cycle from state $|0\rangle$ to $|1\rangle$ and then back to $|0
\rangle$ with a $\pi$ phase shift in the amplitude of $|0\rangle$. Therefore the leakage amplitude accumulated during the next four $\pi/2$ pulses will have an additional $-1$ factor. After $2N$ repetitions ($n_{\rm reps}=2N$), assuming $N$ is not too large such that leakage amplitude in $|2\rangle$ is still relatively small and we can neglect state $|2\rangle$ amplitude coming back to the computational subspace, the total amplitude in state $|2\rangle$ can be approximated as
\begin{align}
\label{eq:c2_nreps=2n}
c_2(n_{\rm reps}=2N) &= c_2(n_{\rm reps}=2)e^{i2\pi (N-1) \eta t_{\rm cyc}}\nonumber\\
&\times \sum_{m=0}^{m=N-1}(-1)^m e^{-i2\pi m \eta t_{\rm cyc}},\\
t_{\rm cyc}&\equiv  2 (2t_{\rm gate} + t_{\rm delay}).
\end{align}
For simplicity of notation, we introduced cycle time $t_{\rm cyc}$ to denote the period of two repetitions.

In order to have constructive interference between every term in the right-hand side of Eq.~(\ref{eq:c2_nreps=2n}), we need to satisfy the following condition:
\begin{align}
    \eta  t_{\rm cyc} = k + 1/2, \quad k=0,1,2,...
\end{align}
This is the condition for the locations of peaks in Fig.~2(b) of the main text. Under this condition, we have $|c_2(n_{\rm reps}=2N)|=N|c_2(n_{\rm reps}=2)|$. Height of peaks can be different for even and odd $k$ because the magnitude of $c_2(n_{\rm reps}=2)$ is generally different between them. It follows from Eq.~(\ref{eq:c2_nreps=2}) that, for even $k$ (corresponding to type-A peaks in Fig.~2 of the main text, we have $c_2(n_{\rm reps}=2) = ic + c'$, for odd $k$ (corresponding to type-B peaks), we have $c_2(n_{\rm reps}=2) = -ic + c'$. Then it follows that heights of A and B peaks are approximately given by the following ($n_{\rm reps} = 2N$):
\begin{align}
\label{eq:a_peak_height}
    p_2({\rm A~peak})= N^2 |c-ic'|^2 = N^2 [|c|^2 + |c'|^2 - 2{\rm Im}(cc'^*)], \\
\label{eq:b_peak_height}
    p_2({\rm B~peak})= N^2 |c+ic'|^2 = N^2 [|c|^2 + |c'|^2 + 2{\rm Im}(cc'^*)].
\end{align}
Recall that $c$ and $c'$ are the amplitudes accumulated in state $|2\rangle$ when qubit goes from state $|0\rangle$ to $|1\rangle$ after two $\pi/2$ pulses and from $|1\rangle$ to $|0\rangle$ after another two $\pi/2$ pulses, respectively. In general, we do not expect the phase difference between these amplitudes to be exactly 0 or $\pi$, therefore heights of A and B peaks are generally different according to Eqs.~(\ref{eq:a_peak_height},\ref{eq:b_peak_height}).

\section{Additional numerical simulation results}
\label{app:additional_simulation}

\begin{figure}[t]
    \centering
    \includegraphics[width = 0.24\textwidth]{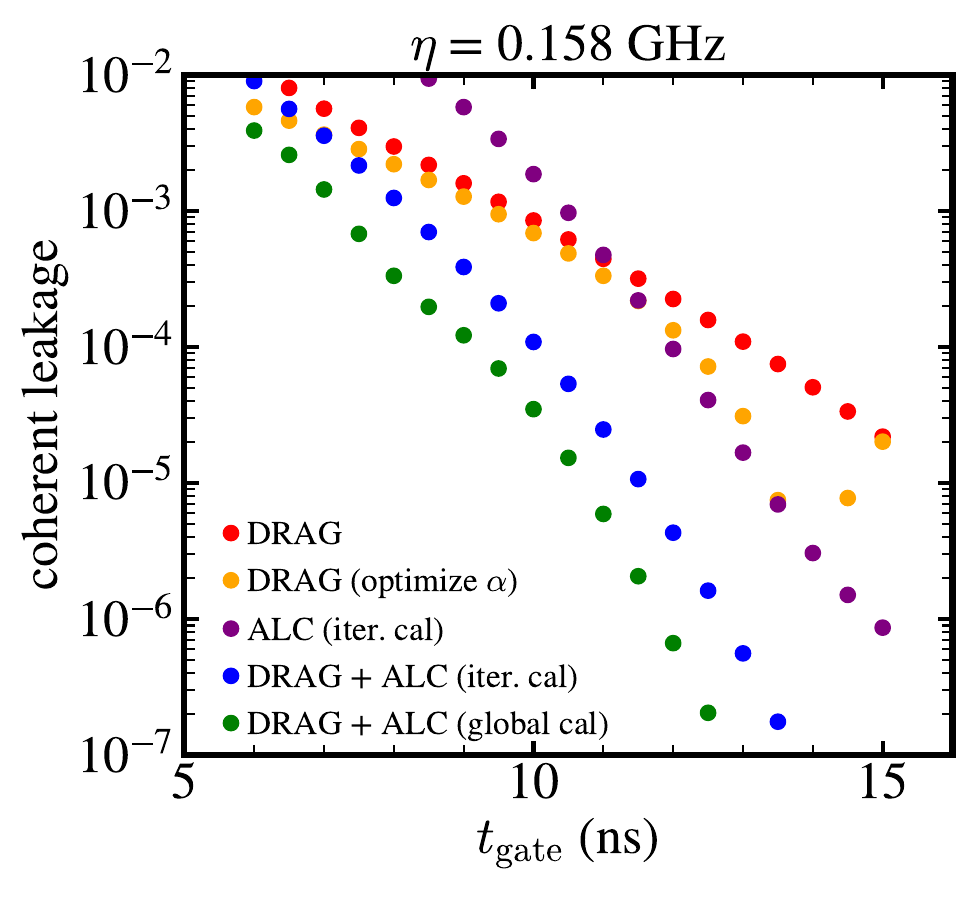}\hfill
    \includegraphics[width = 0.24\textwidth]{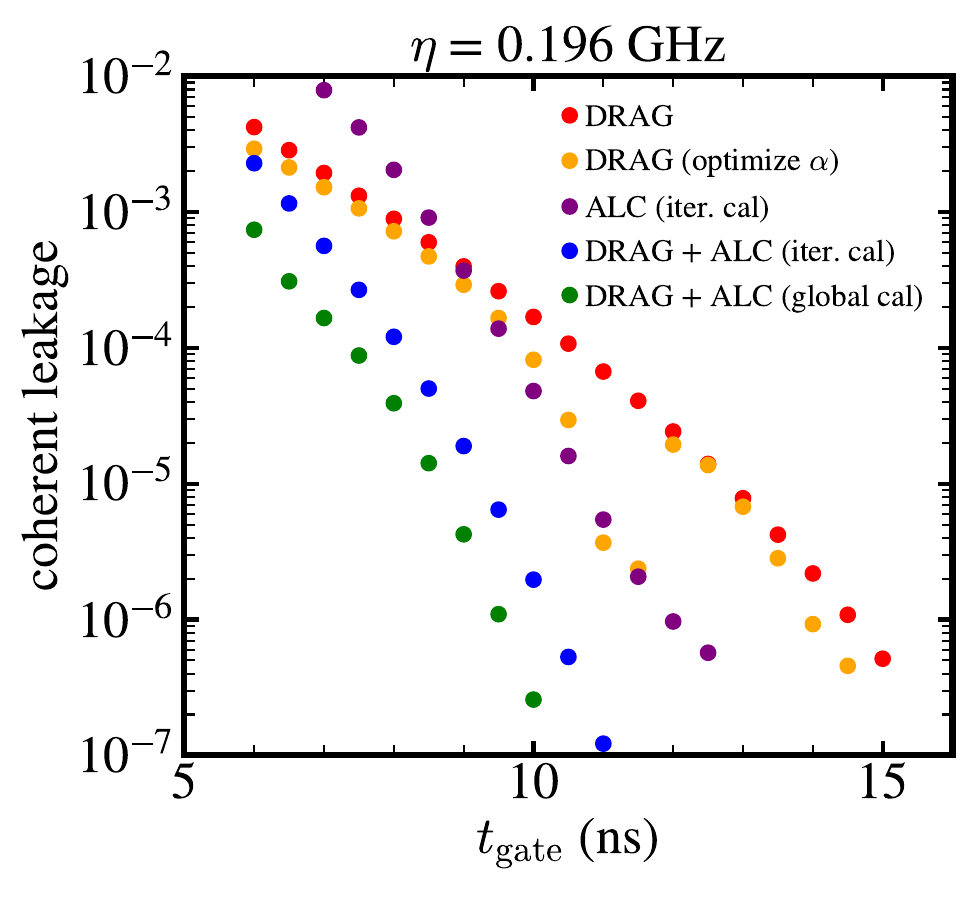}
    \caption{Numerical simulations of coherent leakage after a $\pi/2$ pulse for $\eta = 0.158$ GHz (left) and 0.196~GHz (right). Markers of different colors refer to different leakage suppression strategy: red markers refer to applying only DRAG ($\alpha = 1$), orange markers refer to applying DRAG with optimized DRAG parameter $\alpha$, purple markers refer to applying ALC drive but without DRAG for the main pulse (i.e., $\alpha = 0$), blue and green markers refer to combining DRAG ($\alpha = 1$) and ALC drive but with different optimization procedure (iterative vs global); see the text for details. The leakage shown here only includes drive-induced coherent leakage into states outside of the computational subspace, and does not include additional background leakage due to heating as shown in Fig.~3(a,b) of the main text. Blue and green markers greater than 13~ns (10~ns) on the left (right) panel are lower than $10^{-7}$. }
    \label{fig:global_cal}
\end{figure}
In this section, we show additional numerical simulation results based on the Hamiltonian in Eq.~(1) of the main text. Four transmon states are included in the simulations.

\begin{figure}[t]
    \centering
    \includegraphics[width = 0.47\textwidth]{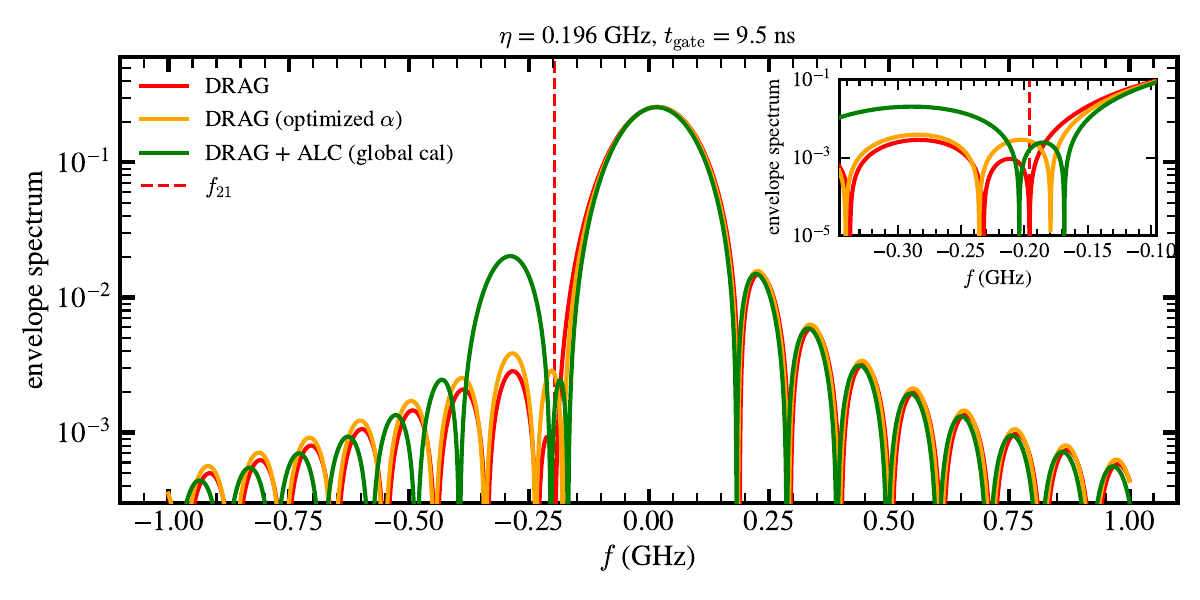}\\
    \includegraphics[width = 0.47\textwidth]{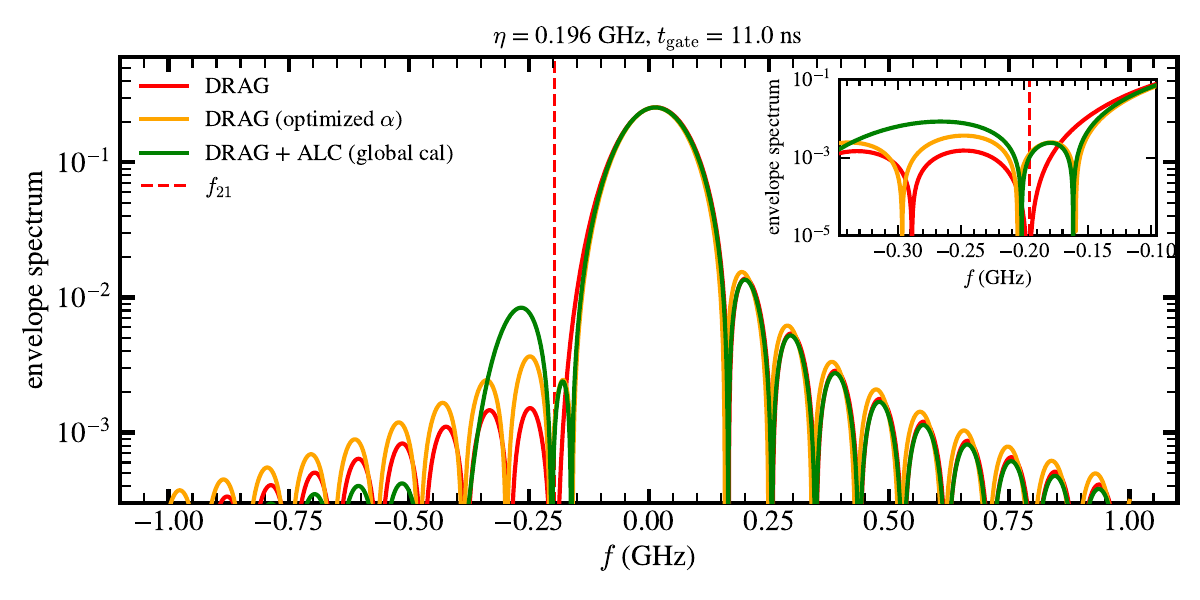}
    \caption{Spectra of pulse envelopes of different leakage suppression strategy  corresponding to $\eta = 0.196$~GHz (Fig.~\ref{fig:global_cal} right panel) at two different gate times: $t_{\rm gate} = 9.5$~ns (top), $t_{\rm gate}$ = 11~ns (bottom). The spectra are centered at qubit $f_{10}$. Same color encoding as Fig.~\ref{fig:global_cal} are used to indicate different leakage suppression strategy. Insets show the zoomed-in spectra near $f_{21}$. }
    \label{fig:spectra}
\end{figure}

In Fig.~\ref{fig:global_cal}, we compare coherent leakage for a $\pi/2$ pulse between different leakage suppression strategies. Coherent leakage includes the sum of final populations in states $|2\rangle$ and $|3\rangle$ at the end of the gate.  Red and orange markers refer to using DRAG only with fixed DRAG parameter $\alpha = 1$ and optimized DRAG parameter, respectively. Purple markers refer to applying ALC drive without DRAG using iterative optimization. Blue and green markers refer to combining DRAG (with $\alpha$ fixed at 1) and ALC using iterative optimization and global optimization, respectively.  Iterative optimization follows the same three-step procedure as used in the experiments: in step 1, we optimize the primary pulse parameters to minimize the computational space error in the absence of ALC drive; in step 2, we optimize ALC drive parameters to minimize leakage to state $\ket{2}$; in step 3, we re-optimize primary drive parameters in the presence of ALC drive. In the global optimization, we optimize pulse parameters all together to minimize the total gate error defined as the squared state overlap between the target and actual state averaged over six different initial states, $\ket{0},\ket{1},\ket{0}\pm\ket{1},\ket{0}\pm i\ket{1}$. In all the five cases, we verified numerically that the remaining gate error after the optimization is dominated by the coherent leakage. In all cases but DRAG + ALC (global cal.), residual coherent leakage mainly consists of leakage to state $|2\rangle$. In the case of DRAG + ALC (global cal.), at long gate times ($t_{\rm gate}\gtrsim 2/\eta$), coherent leakage is dominated by leakage to state $|3\rangle$. As gate time decreases, leakage to state $|2\rangle$ starts to increase and it becomes comparable to state $|3\rangle$ leakage for $t_{\rm gate}\lesssim 1.4/\eta$.

Figure~\ref{fig:global_cal} demonstrates substantial coherent leakage reduction by using DRAG + ALC compared with using DRAG only. Furthermore, DRAG + ALC with globally optimized primary and ALC drive parameters (green) outperforms the iterative optimization (blue) by 2 to 8 fold. Lastly, we show that simply optimizing the drag parameter $\alpha$ (orange) in the standard DRAG scheme does not achieve the same performance as DRAG + ALC. While optimizing DRAG parameter $\alpha$ reduces leakage compared with fixed $\alpha$ at certain gate times, the improvement at other gate times are marginal. Even for the gate times that show improvement, the improvement are significantly less than using DRAG + ALC.

Figure~\ref{fig:spectra} compares the spectra of pulse envelopes of different leakage suppression strategies. The standard DRAG with $\alpha = 1$ has a notch exactly at frequency $f_{21}$. In contrast, DRAG with optimized $\alpha$ and DRAG + ALC have notches at frequencies slightly shifted from $f_{21}$. Interestingly, in the case in which DRAG with optimized $\alpha$ shows significant improvement compared with DRAG with $\alpha$ fixed at 1, the locations of the notches almost match with those of DRAG + ALC; see the green and orange curves in Fig.~\ref{fig:spectra} inset. Yet even in this case, DRAG + ALC shows substantial leakage reduction compared to DRAG with optimized $\alpha$. This indicates that to achieve the very strong leakage suppression as shown in DRAG + ALC, simply creating spectral notches near $f_{21}$ is not sufficient; rather, the ALC drive works by utilizing destructive interference in the leakage amplitudes through fine tuning its frequency and amplitude; see Appendix~B of the main text.

A comment is in order. In Fig.~\ref{fig:global_cal} left panel, the leakage suppression performance of DRAG + ALC (iterative cal) is better than those shown in Fig.~3 of the main text for the same anharmonicity of $\eta = 158$~MHz. This is because after the experimental data in Fig.~3(a,c) of the main text were collected, we found that using $\Delta = -(\eta + \delta_{\rm main})$ for the primary pulse [see Eq.~(2) of the main text] achieves better performance in leakage cancellation (about a factor of 3 to 4 further reduction in leakage) than using $\Delta = -\eta$, which was used in Fig.~3(a,c) of the main text. In Fig.~\ref{fig:global_cal}, we show the better results using $\Delta = -(\eta + \delta_{\rm main})$. In all the experimental and theoretical results for $\eta = $ 196~MHz, we use $\Delta = -(\eta + \delta_{\rm main})$.

\section{Robustness against small shift in qubit frequency}

\begin{figure}[t]
    \centering
    \includegraphics[width = 0.24\textwidth]{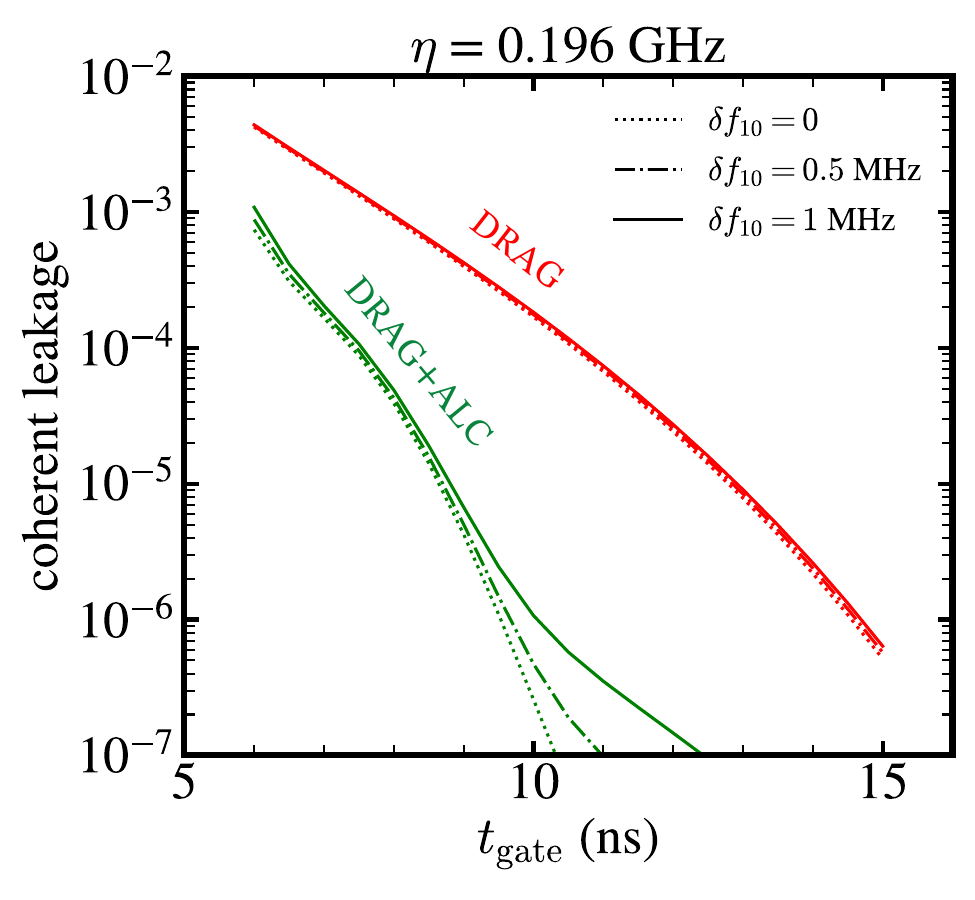}\hfill
    \includegraphics[width = 0.24\textwidth]{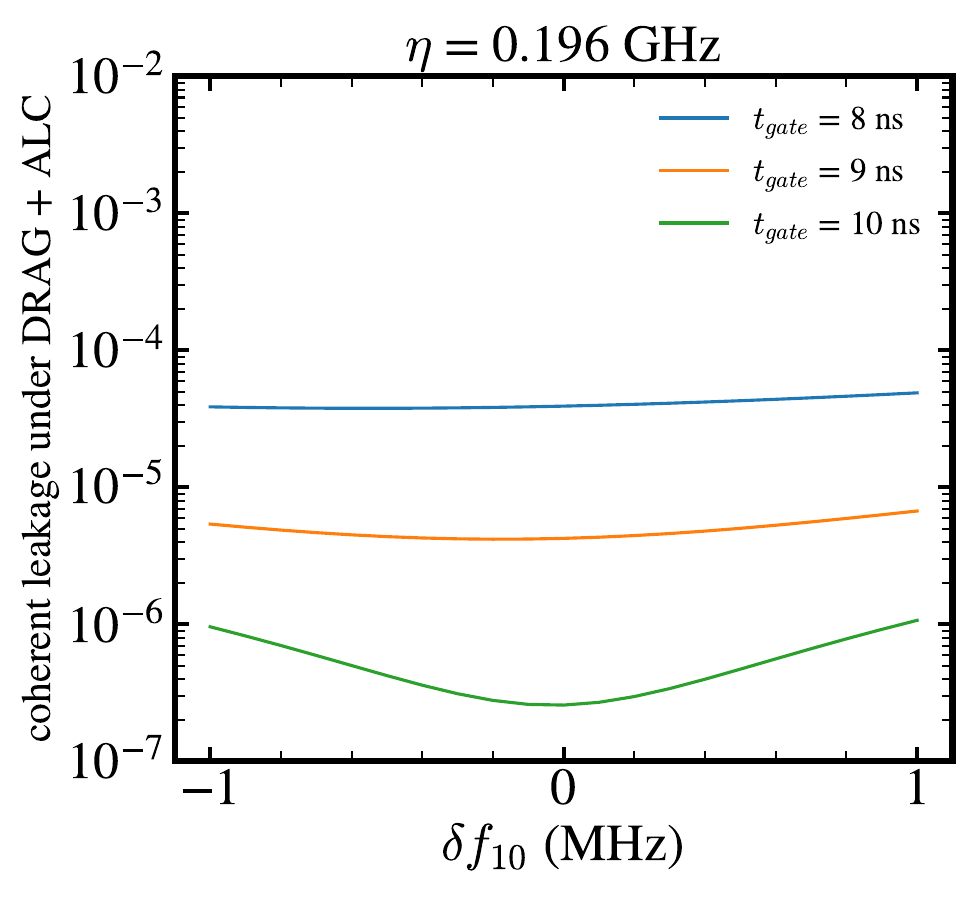}
    \caption{Robustness of leakage cancellation against static qubit frequency shift. Left panel: compare coherent leakage at the end of a $\pi/2$ pulse between DRAG shown in red and DRAG + ALC (global cal.) shown in green for various gate times in the presence of small qubit frequency shift $\delta f_{10}$. Different line styles refer to different frequency shift: $\delta f_{10}$ = 0 (dotted), 0.5 MHz (dash dotted), 1 MHz (solid). Right panel: coherent leakage under DRAG + ALC for various qubit frequency shifts at fixed gate times. Different colors refer to different gate times: $t_{\rm gate}$ = 8 ns (blue), 9 ns (orange), 10 ns (green). The flatter the curve is, the more robust the leakage cancellation against qubit frequency shift is. Minimum of the coherent leakage is not exactly at $\delta f_{10} = 0$ because the numerical simulation finds pulse parameters that minimize the total gate error which includes both coherent leakage and errors that occur in the computational subspace.}
    \label{fig:freq_shift_robustness}
\end{figure}

In this section, we numerically study the robustness of leakage cancellation performance against small shift in qubit frequency. For transmon qubits, qubit frequency shifts can come from residual ZZ coupling with nearby qubits which typically range from tens to hundreds of kHz; see for instance Ref.~[22]. For flux tunable qubits, they can also come from low-frequency flux noise such as 1/f noise~[23].

To simulate the effect of qubit frequency shift, we first find optimal pulse parameters in the absence of qubit frequency shift according to the procedure described in Sec.~\ref{app:additional_simulation}. Then we introduce a small static shift $\delta f_{10}$ in qubit frequency while keeping qubit anharmonicity fixed; this means both qubit $f_{10}$ and $f_{21}$ shift by the same amount. This shifts the drive detuning parameters ($\delta_{\rm main}$ for the main drive and $\delta_{\rm alc}$ for the ALC drive) away from the optimized values. We then compute the coherent leakage at the end of the gate using these shifted drive parameters by simulating the Hamiltonian in Eq.~(1) of the main text. 

Figure~\ref{fig:freq_shift_robustness} shows the simulated coherent leakage at the end of a $\pi/2$ pulse in the presence of qubit frequency shifts. The results show that even for a qubit frequency shift as large as 1~MHz, the strong leakage suppression of DRAG + ALC compared to DRAG remains essentially unchanged. At longer gate times (above 10 ns), coherent leakage under DRAG + ALC increases somewhat significantly with $\delta f_{10}$, but it remains to be several order of magnitude smaller than using standard DRAG. As gate time becomes shorter, performance of leakage cancellation becomes increasingly more robust with respect to $\delta f_{10}$ as indicated by the flatness of the curves in the right panel. This makes sense intuitively because as the pulse becomes shorter, it becomes larger in strength and therefore more robust against small change in detuning. 

\section{Discussion on crosstalk and ALC drive}
Experiments shown in this paper were performed for isolated qubits. In the presence of surrounding qubits, microwave drive on one qubit can potentially spread to another qubit, known as microwave crosstalk. The additional crosstalk due to the ALC drive can in principle be compensated using the mitigation technique described in the supplementary material of Refs.~[1,2]. Given that the ALC drive is typically much weaker than the main drive, we anticipate the crosstalk due to the ALC drive to be a weaker effect than that from the main drive. Detailed analysis and demonstration is beyond the scope of this paper.

\end{document}